\documentclass[journal]{IEEEtran}

\usepackage{graphicx}
\usepackage{float}
\usepackage{mathrsfs}
\usepackage{amsfonts}
\usepackage{subeqn}

\usepackage{cite}

\usepackage{booktabs} %for san xian biao

\ifCLASSINFOpdf

\else

\fi

\begin{document}

\title{Joint Multiple Symbol Differential Detection and Channel Decoding for Noncoherent UWB Impulse Radio by Belief Propagation}

\author{Taotao Wang, Tiejun Lv, \emph{Senior Member, IEEE,} Hui Gao, \emph{Member, IEEE}, Shengli  Zhang, \emph{Member, IEEE}% <-this % stops a space

\thanks{T. Wang is with the College of Information Engineering, Shenzhen University, China, and the Department of Information Engineering, The Chinese University of Hong Kong, Hong Kong (email: wtt011@ie.cuhk.edu.hk). }
\thanks{T. Lv and H. Gao are with the School of Information and Communication Engineering, Beijing University of Posts and Telecommunications, China (email: \{lvtiejun, huigao\}@bupt.edu.cn). }
\thanks{S. Zhang is with the College of Information Engineering, Shenzhen University, China (email: zsl@szu.edu.cn).}
%\thanks{T. Wang is supported by the Hong Kong PhD Fellowship Scheme. T. Lv is supported by the National Natural Science Foundation of China (NSFC) (Grant No. 61271188) and the China 973 Program  (No. 2011CB302702).}
}
% The paper headers
%\markboth{This work is submitted to IEEE Transactions on Wireless Communication for peer review}%
%\markboth{Joint Noncoherent Detection and Channel Decoding for UWB Impulse Radio by Belief Propagation}%
%{Shell \MakeLowercase{\textit{et al.}}: Bare Demo of IEEEtran.cls
%for Journals}

% make the title area
\maketitle

\begin{abstract}
%\boldmath
This paper proposes a belief propagation (BP) message passing algorithm based joint multiple symbol differential detection (MSDD) and channel decoding scheme for noncoherent differential ultra-wideband impulse radio (UWB-IR) systems. MSDD is an effective means to improve the performance of noncoherent differential UWB-IR systems. To optimize the overall detection and decoding performance, in this paper, we propose a novel soft-in soft-out (SISO) MSDD scheme and its integration with SISO channel decoding for noncoherent differential UWB-IR. we first propose a new auto-correlation receiver (AcR) architecture to sample the received UWB-IR signal. The proposed AcR can exploit the dependencies (imposed by the differential modulation) among data symbols throughout the whole packet. The signal probabilistic model has a hidden Markov chain structure. We use a factor graph to represent this hidden Markov chain. Then, we apply BP message passing algorithm on the factor graph to develop a SISO MSDD scheme, which has better performance than the previous MSDD scheme and is easy to be integrated with SISO channel decoding to form a joint MSDD and channel decoding scheme. Simulation results indicate the performance advantages of our MSDD scheme and joint MSDD and channel decoding scheme.
\end{abstract}


\IEEEpeerreviewmaketitle

\section{Introduction}
% no \IEEEPARstart
\IEEEPARstart{U}{ltra}-wideband impulse radio (UWB-IR) is served as a promising candidate for location-aware indoor communications, wireless sensor networks and wireless personal area networks. Previously, UWB-IR earned significant attentions in both academia and industry \cite{yang2004uwc}. However, the implementation of optimal coherent receiver for UWB-IR systems faces many challenges. UWB channels usually contain hundreds of multipath, due to the rich scattering indoor environments. The optimal coherent receiver required to capture multipath energy is the famous Rake receiver \cite{rajeswaran2003rake}. Since the UWB channel is characterized by the dense multipath, we need a large number of Rake fingers to capture a significant part of the signal energy \cite{win1998ecu}. The implementation of so many Rake fingers and the associated channel estimation on the corresponding multipaths involve intensive complexities \cite{lottici2002channelestimation}. Moreover, such Rake receiver is very sensitive to timing-jitter \cite{chen2006timing}. These challenges make it difficult and costly to realize the optimal coherent receiver for UWB-IR systems.

To obviate the complicated treatments on UWB channels, the suboptimal noncoherent receivers are proposed \cite{witrisal2009noncoherent}. The typical noncoherent UWB-IR schemes are differential \cite{ho2002differential} and transmitted-reference \cite{choi2002performance} UWB-IR systems, both deployed with the analog autocorrelation receiver (AcR) that does not require Rake receiver and explicit channel estimation.  Due to their good performance-complexity tradeoff, noncoherent receivers are now more popularly used in UWB-IR systems. However, they suffer from some performance degradations compared with coherent receivers.

Multiple symbol differential detection (MSDD) is an effective means to improve
the performance of noncoherent differential UWB-IR systems. The theoretical framework
of MSDD is the maximum-likelihood (ML) sequence detection, which is
firstly introduced to detect a block of differential MPSK symbols over additive
white Gaussian noise (AWGN) channel in \cite{divsalar1990multiple}.  Applying MSDD to differential UWB-IR systems is considered in \cite{guo2006improved,
yang2008noncoherent, lottici2008multiple}.  Works \cite{lottici2008multiple, wang2011sphere} consider the application of sphere decoding algorithm to fulfill a low complexity MSDD for differential UWB-IR systems. With the same purpose, work \cite{lottici2008multiple} proposed a Viterbi algorithm based MSDD scheme and works \cite{qi2010fast, schenk2011decision, wang2013ber} proposed decision-feedback MSDD schemes.

Wireless communication systems are susceptible to various impairments, such as noises, interferences and channel fading. Channel codes are usually employed to protect the transmitted symbols over possible errors. The decoding of most powerful channel codes that can approach the Shanon capacity depends on iterative algorithm, where iterations are performed between soft-in soft-out (SISO) modules \cite{wiberg1995codes}. These MSDD schemes mentioned earlier, however, are all about to detect the hard decisions of the differential modulated UWB-IR signals, which is not compatible with SISO channel decoding. Recently, the work \cite{zhou2012soft} investigates SISO MSDD for UWB-IR systems and incorporate it with SISO channel decoding.

In this paper, we propose a new SISO MSDD scheme for noncoherent differential UWB-IR systems. Even without considering channel encoding, there are memories introduced by the differential modulation to all modulated symbols throughout the whole packet. In \cite{zhou2012soft}, the SISO MSDD processes signal samples block-by-block and it just ignores the information dependencies among different blocks. This leads to information loss. In this paper, in contrast, the proposed SISO MSDD scheme calculates the soft information of one symbol by exploiting the signal samples from the whole packet. We propose an AcR architecture to enable this scheme. The proposed AcR architecture correlates the received UWB-IR signals, and does not need explicit channel estimation. Thus, the proposed AcR is a noncoherent receiver. Moreover, it can exploit the signal dependencies among the whole packet. The joint probability function of the data symbols and the correlation samples in the packet has a hidden Markov chain structure. We use a factor graph to represent this hidden Markov chain of the signal probabilistic model. We then develop our SISO MSDD scheme using the belief propagation (BP) message passing algorithm which implements sum-product rule on the factor graph \cite{kschischang2001factor}. The proposed MSDD scheme has better detection performance than the previous block-by-block MSDD scheme in \cite{guo2006improved, yang2008noncoherent, lottici2008multiple}. We also consider the channel decoding for noncoherent differential UWB-IR systems. Since BP message passing algorithm is also employed as the decoding algorithm for many channel codes, we integrate the proposed MSDD and the BP messag passing algorithm for SISO channel decoding. The outputs of the MSDD are fed to the inputs of the channel decoding, and vice versa, in an iterative manner. The main contributions of this paper are summarized as the follows.

\begin{enumerate}

\item \emph{New noncoherent AcR architecture for generating correlation samples.} We propose a new noncoherent AcR architecture to sample the received UWB-IR signal. Compared with the existing AcR in \cite{guo2006improved, yang2008noncoherent, lottici2008multiple}, the proposed AcR can exploit more signal dependencies imposed by the differential modulation. The proposed AcR results in a hidden Markov chain model for the signal of the whole packet.

\item \emph{BP message passing algorithm for SISO MSDD scheme (joint MSDD and channel decoding scheme).} We apply the BP message passing algorithm to the factor graph that represents the hidden-Markov-chain-type signal model for deriving SISO MSDD scheme. The proposed MSDD scheme is a bidirectional algorithm that consists of a forward and a backward message passing. Furthermore, we integrate the proposed MSDD with the BP messag passing algorithm for SISO channel decoding, and we achieve an iterative algorithm for the joint MSDD and channel decoding scheme. We believe this is the first time that applies BP message passing algorithm to noncoherent differential UWB-IR systems.

\item \emph{Performance evaluations by simulations.}  Simulations are performed to validate and evaluate the proposed scheme. The performances of uncoded and coded system under the environments of UWB multipath channel are evaluated. The results indicate the performance advantage of the proposed scheme over other existing schemes.

\end{enumerate}

The rest of this paper is organized as follows. The system model of
differential UWB-IR system is described in Section II. Section III introduces the joint MSDD and channel decoding scheme. Section IV shows simulation results. Finally, conclusions are drawn in Section V.

\section{System Model}
\begin{figure}[!t]
	\centering
	\includegraphics[width=3.5in]{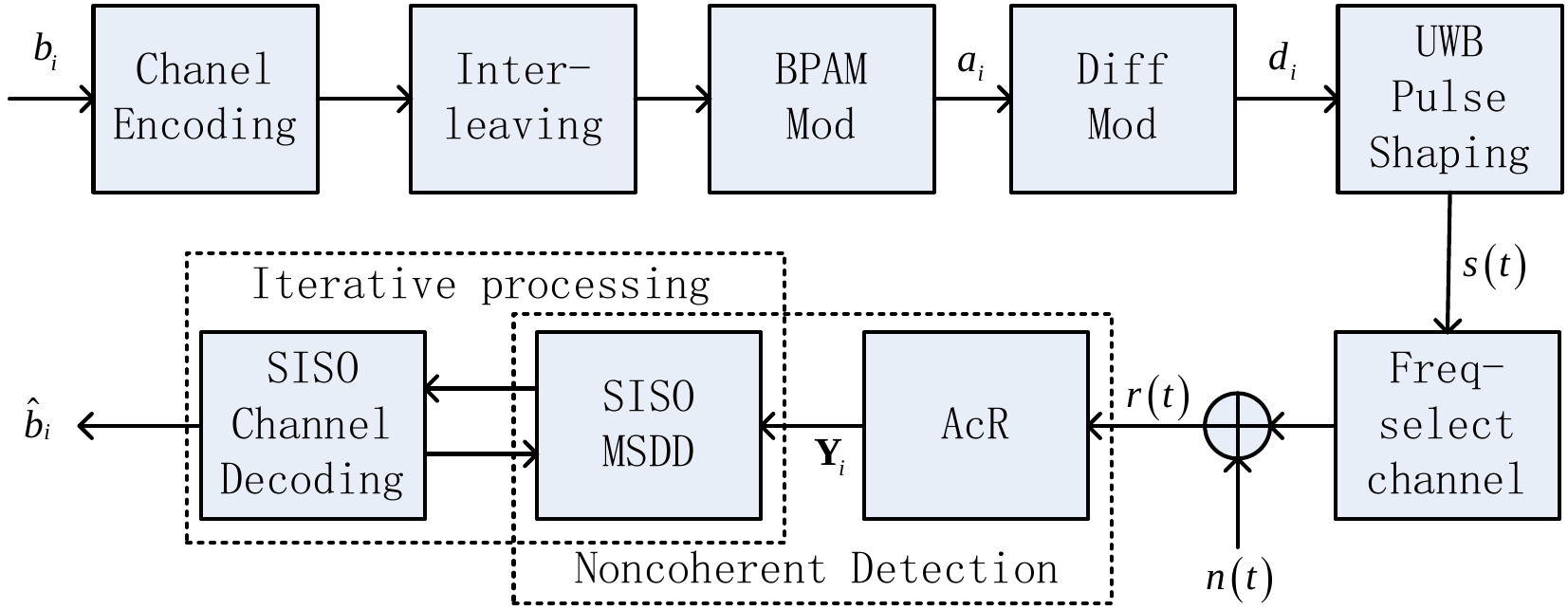}
	\caption{A block schematic diagram of the system model.} \label{sysmod}
\end{figure}

In this section, we present the system model for UWB-IR communications.  A block schematic diagram of the system model is shown in Fig.~\ref{sysmod}.  Adopting binary antipodal pulse amplitude modulation (BPAM), the transmitted signal waveform is given by
\begin{equation}\label{tx_signal_1}
s\left( t \right) = \sum\limits_{i = 0}^{N} {{d_i}{\omega _s}\left( {t - i{T_s}} \right)},
\end{equation}
where ${{d_i}} \in \left\{ { \pm 1} \right\}$ is the $i^{th}$ channel symbol, ${\omega _s}\left( t \right)$ is the symbol waveform with duration $T_s$ and $N$ is the packet size.  We denote the original information bits by ${b_j} \in \left\{ {0,1} \right\}$, $j = 1, \cdots ,K$. Through channel encoding, interleaving and modulation, these information bits are mapped to the coded data symbols ${a_i} \in \left\{ {\pm 1} \right\}$, $i = 1, \cdots ,N$. The coding rate is $R = {K \mathord{\left/{\vphantom {K N}} \right. \kern-\nulldelimiterspace} N}$. Finally, the channel symbol ${{d_i}}$ is obtained by differential modulation: ${d_i} = d_{i - 1}a_i$, $i = 1, \cdots ,N$, where ${d_0} = 1$ is the reference symbol. UWB-IR transmissions usually employ ${N_f}$ frames to transmit one channel symbol, and each frame includes one very short pulse. According to this unique aspect of UWB-IR, the unmodulated symbol waveform used in (\ref{tx_signal_1}) is expressed as
\begin{equation}\label{sym_pulse}
{\omega _s}\left( t \right) = \sum\limits_{j = 0}^{{N_f} - 1} {{\omega}\left( {t - j{T_f} - {c_j}{T_c}} \right)},
\end{equation}
where $\omega \left( t \right)$ is the ultra-short pulse with the duration ${T_\omega }$ (referred to as the monocycle in literatures), $T_f$ is the frame duration and we have $T_s=N_fT_f$.  The sequence $\left\{ {{c_j}} \right\}$ in (\ref{sym_pulse}) is a user specific time-hopping (TH) code used for the purpose of multiple access. Its elements are integers in the range $0 \le {c_j} \le {N_c} - 1$, satisfying ${T_f} \ge {N_c}{T_c}$.  ${T_c}$ is the duration of an addressable time chip. Since $\omega \left( t \right)$ has a very short duration, $T_\omega$ is typically on the order of nanoseconds, the transmitted signal occupies a huge bandwidth. The frame duration  $T_f$ is usually hundred or thousand times longer than $T_\omega$, resulting
in a low duty transmission.

We consider dense multi-path environments, such as the industrial and indoor office \cite{molisch2006comprehensive}. The channel impulse response (CIR) between the transmitter and the receiver is modeled as $h\left( t \right) = \sum\nolimits_{l = 0}^{L - 1} {{\alpha _l}\delta \left( {t - {\tau _l}} \right)}$, where $\delta$ is the Dirac delta function, $L$ is the number of resolvable multipath components (MPCs), $ \alpha _l$ and $\tau _l$ is the gain and the delay of the $l^{th}$ MPC, respectively.

We define the received pulse waveform as: $g\left( t \right) \buildrel \Delta \over = \omega \left( t \right) \otimes h\left( t \right)$, where $ \otimes$ denotes the convolution operator.
Then, the received noisy signal waveform is given by
\begin{equation}\label{rx_siganl}
\begin{array}{l}
 r\left( t \right) = s\left( t \right) \otimes h\left( t \right) + n\left( t \right) \\
  \;\;\;\;\;\;\;\:= \sum\limits_{i = 0}^N {{d_i}\sum\limits_{j = 0}^{{N_f} - 1} {g\left( {t - i{T_s} - j{T_f} - {c_j}{T_c}} \right)} }  + n\left( t \right), \\
 \end{array}
\end{equation}
where $n\left( t \right)$ is the additive white Gaussian noise process with zero mean and a flat two-sided power spectral density $ {{N_0 } \mathord{\left/{\vphantom {{N_0 } 2}} \right. \kern-\nulldelimiterspace} 2}$. With $T_g$ denoting the maximum delay spread of the received pulse waveform  $g\left( t \right)$, the inter-frame interference (IFI) is avoided in the received signal (\ref{rx_siganl}) by letting $T_f  > T_g  + \left( {{N_c} - 1} \right)T_c$.

\section{Joint Noncoherent Detection and Channel Decoding}

\subsection{New Noncoherent Autocorrelation Receiver}

The optimal coherent detection of UWB-IR signals requires an implementation of the filter matched to the received pulse waveform $g\left( t \right)$. However, the complexities of the implementation of the match filtering and the explicit channel estimation constitute obstacles for the practical use of coherent detection in UWB-IR systems. Therefore, we focus on noncoherent detection that dose not involve the explicit channel estimation and the implementation of the match filtering. To improve the performance of noncoherent detection, we apply the MSDD scheme to the system. UWB channels are quasi-static in typical indoor environments \cite{molisch2006comprehensive}. This means the channel remains invariant over several symbol durations. Relying on this feature, MSDD jointly detects a block of $M$ symbols from the received signal in the observation window of size $M + 1$ symbol durations \cite{guo2006improved, lottici2008multiple, zhou2012soft}.

\begin{figure}[!t]
\centering
\includegraphics[width=3.5in]{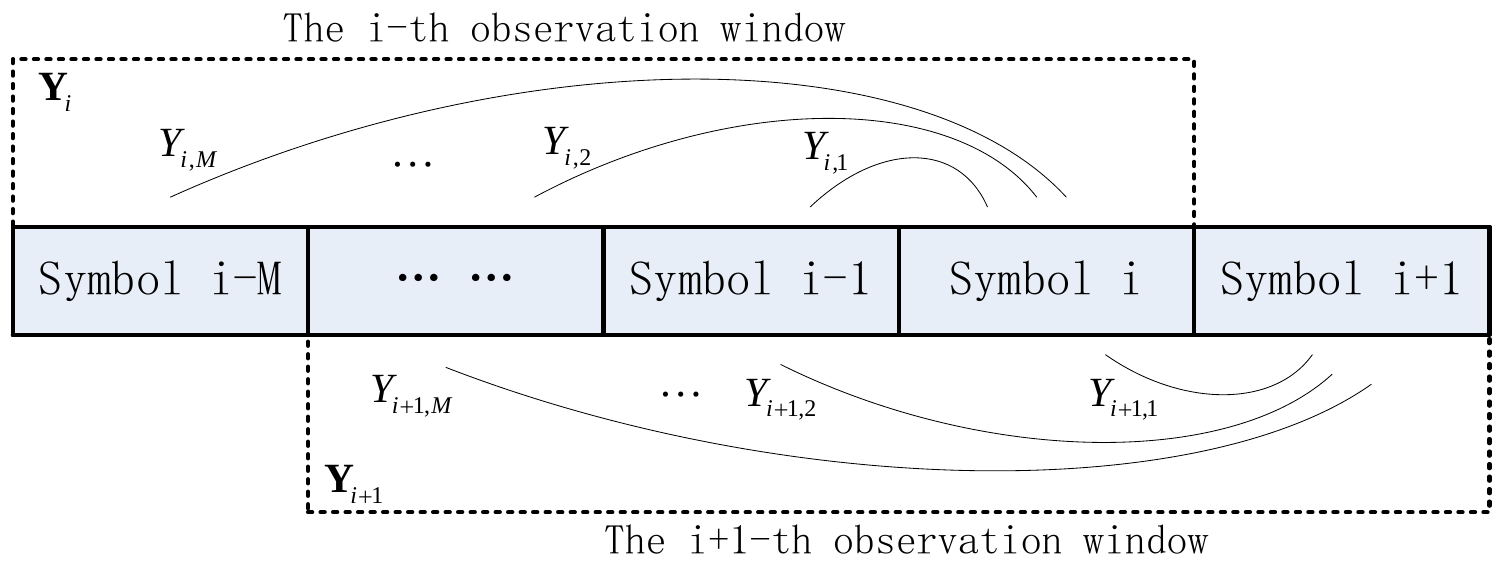}
\caption{The illustration for the sampling mechanism of the proposed AcR.} \label{sampling}
\end{figure}

In this section, based on the concept of MSDD, we develop an AcR architecture for noncoherent detection of UWB-IR signals. We modify the sampling mechanism of the MSDD in \cite{guo2006improved, lottici2008multiple, zhou2012soft}.  Essentially, we still employ the correlation principle derived from GLRT criteria in \cite{lottici2008multiple} to sample the received signal; however, we change its sliding mode of the observation window. In \cite{lottici2008multiple}, each time, the observation window of size $M+1$ will be slid down $M$ symbol durations after the current $M$ symbols have been detected. In a different mode, we slide the $M+1$ size  observation window down one symbol duration each time. In other words, the current observation window overlaps $M$ symbols with the next  observation window. The sampling mechanism of the proposed AcR is illustrated in Fig. \ref{sampling}.  In the following, we mathematically formulate the proposed sampling mechanism, and then explain its implications to noncoherent UWB-IR systems.

From the received signal $r\left( t \right)$ in the ${i^{th}}$ observation window $\left( {i - M} \right){T_s} \le t \le \left( {i + 1} \right){T_s}$,  we obtain the ${i^{th}}$  sample vector ${{\bf{Y}}_i} \buildrel \Delta \over = {\left[ {{Y_{i,1}},{Y_{i,2}}, \cdots ,{Y_{i,M}}} \right]^T}$, where ${Y_{i,m}}$ is the correlation sample between the $i^{th}$ and the $(i-m)^{th}$ symbols
\begin{equation}\label{sig_sampling}
\begin{array}{l}
{Y_{i,m}} = \int\limits_0^{{T_g}} {y\left( {t + i{T_s}} \right)y\left( {t + \left( {i - m} \right){T_s}} \right)dt} ,\\ 
\;\;\;\;\;\;\;\;\;\;\;\;\;\;\;\;\;\;\;\;\;\;\;\;\;\;\;\;\;\;\;\;\;\;\;\;\;\;\;\;\;\;\;\;\;\;\;\;\;\;\;  m = 1,2, \cdots, M  \\
\;\;\;\;\;\;\;\;\;\;\;\;\;\;\;\;\;\;\;\;\;\;\;\;\;\;\;\;\;\;\;\;\;\;\;\;\;\;\;\;\;\;\;\;\;\;\;\;\;\;\;  i = 1,2, \cdots, N 
\end{array}
\end{equation}
with the de-spreading signal
\begin{equation}\label{despreading_signal}
y\left( t \right) = \sum\limits_{j = 0}^{{N_f} - 1} {r\left( {t + j{T_f} + {c_j}{T_c}} \right)}.
\end{equation}
After we finish the computation of ${{\bf{Y}}_i}$,  the observation window is slid down one symbol duration to $\left( {i +1- M} \right){T_s} \le t \le \left( {i + 2} \right){T_s}$, from where we will compute the next sample vector ${{\bf{Y}}_{i+1}}$.  Since there is no transmission occurring ($r\left( t \right) = 0$ for $t < 0$), we pad some zeros at the rears of the first $M-1$ sample vectors: ${Y_{i,m}} = 0$  for $i = 1, \cdots ,M - 1$ and $m = i + 1, \cdots ,M$.  We make some remarks about the proposed AcR to bring out its implications.

\begin{enumerate}

\item   The AcR receiver can be realized using analog components to avoid the ultra high sampling rate in UWB-IR systems. The sampling mechanism of the proposed AcR produces the samples in ${{\bf{Y}}_i}$ by correlating the $i^{th}$ symbol with its previous $M$ symbols as expressed in (\ref{sig_sampling}); then it slides the observation window down one symbol duration to produce samples in ${{\bf{Y}}_{i+1}}$; the correlating operations are carried out from the first symbol to the last symbol of the packet. By this manner, the proposed AcR can exploit the dependencies (imposed by the differential modulation) among symbols throughout the whole packet.

\item  Based on the samples of the whole packet ${\bf{Y}}_i$ for $i = 1, \cdots ,N$ produced by the proposed AcR, we will derive the BP message passing algorithm on a factor graph for SISO MSDD in the next subsection. We employ a factor graph to represent the probabilistic model of the samples and the data symbols in the whole packet. Then, we apply the BP message passing algorithm to the factor graph for SISO MSDD.  We also combine the BP message passing algorithm for SISO channel decoding with the proposed BP message passing algorithm for SISO MSDD, resulting in a BP message passing algorithm to joint MSDD and channel decoding for UWB-IR systems. Iterative message exchange will be performed between SISO MSDD and SISO channel decoding. A block schematic diagram about this receiver structure is also shown in Fig. \ref{sysmod}.

\item Different from ours, the sampling mechanism of \cite{guo2006improved,
	yang2008noncoherent, lottici2008multiple, zhou2012soft} is on a block-by-block basis. The correlation operations only try to exploit the dependencies (imposed by the differential modulation) among symbols within a block of $M$ symbols. However, the symbol dependencies between different blocks are ignored. This is a kind of information loss. We will discuss the MSDD scheme of \cite{guo2006improved,
	yang2008noncoherent, lottici2008multiple, zhou2012soft} in Section III.D in detail. Depending upon the proposed sampling mechanism in (\ref{sampling}) where blocks overlap some others, the detection of one symbol is capable of making use of the information of all the symbols throughout the whole packet. Since more information are collected, it is expected that the proposed scheme could have better performance.
	
\end{enumerate}

\subsection{The BP Message Passing Algorithm for SISO MSDD}
In this subsection, we derive the BP message passing algorithm for performing SISO MSDD using the samples delivered from the proposed AcR. Here, we do not consider the effect of the channel encoding and we will discuss it in Section III.C.   

Substituting (\ref{despreading_signal}), (\ref{rx_siganl}) into (\ref{sig_sampling}) and using the result of differential demodulation ${d_i}{d_{i - m}} = \prod\nolimits_{z = i- m + 1}^i {{a_z}}$, we can express the sample ${Y_{i,m}}$  as
\begin{equation}\label{dis_sample}
{Y_{i,m}} =\left( {\prod\limits_{z =i- m + 1}^i {{a_z}} } \right) N_f^2{E_g}   + {n_{i,m}},
\end{equation}
where ${E_g} \buildrel \Delta \over = \int_0^{{T_g}} {{g^2}\left( t \right)dt}$ is the captured energy of the received pulse and
\begin{equation}\label{dis_noise}
\begin{array}{l}
 {n_{i,m}} = \int\limits_0^{{T_g}} {y\left( {t + i{T_s}} \right)n\left( {t + \left( {i - m} \right){T_s}} \right)dt}  \\ 
 \;\;\;\;\;\;\;\;\;\;\;\; + \int\limits_0^{{T_g}} {n\left( {t + i{T_s}} \right)y\left( {t + \left( {i - m} \right){T_s}} \right)dt}   \\ 
 \;\;\;\;\;\;\;\;\;\;\;\;  + \int\limits_0^{{T_g}} {n\left( {t + i{T_s}} \right)n\left( {t + \left( {i - m} \right){T_s}} \right)dt}  \\
 \end{array}
\end{equation}
is the discrete noise component. It has been shown in \cite{quek2005analysis} that ${n_{i,m}}$ for all $i$ and $m$ can be approximated to mutually independent Gaussian random variables with mean zero and variance $\sigma _n^2 = {N_f}{N_0}{E_g} + {{W{T_g}N_0^2} \mathord{\left/{\vphantom {{W{T_g}N_0^2} 2}} \right. \kern-\nulldelimiterspace} 2}$, where $W$ is the bandwidth of the bandpass filter employed at the receiver. This approximation is rather well when $N_f$ is large due to the central limit theorem. We will also investigate this Gaussian approximation using simulations in Section IV.   

It can be concluded from (\ref{dis_sample}) that the signal samples depends on the data symbols and the captured energy ${E_g}$. To obtain the knowledge of parameter ${E_g}$, our receiver employs an energy estimation method:
\begin{equation}\label{Eg}
{\hat E_g}{\rm{ = }}\frac{1}{{ZN_f^2}}\sum\limits_{i = 1}^N {\sum\limits_{m = 1}^M {\left| {{Y_{i,m}}} \right|} } 
\end{equation}
where $Z$ is the number of non-zero elements in $\left\{ {{{\bf{Y}}_i}} \right\}$. After that, we detect data symbols by substituting ${\hat E_g}$ into the signal model (\ref{dis_sample}). In the simulation results of Section IV, we will see that this straightforward estimation of  $E_g$ achieves a rather good performance.

Let ${\bf{Y}} \buildrel \Delta \over = \left\{ {{{\bf{Y}}_i}} \right\}$ be the set containing all the sample vectors and ${\bf{a}} \buildrel \Delta \over = {\left[ {{a_1},{a_2}, \cdots ,{a_N}} \right]^T}$ be the vector of all the data symbols. The target of the SISO MSDD is to calculate the \emph{a posteriori} probability (APP) of data symbol $a_i$ given ${\bf{Y}}$:
\begin{equation}\label{app}
p\left( {\left. {{a_i}} \right|{\bf{Y}}} \right) \propto \sum\limits_{{\bf{a}}:\sim {a_i}} {p\left( {\bf{a}} \right)p\left( {\left. {\bf{Y}} \right|{\bf{a}}} \right)} 
\end{equation}
for all $i$, where notation $\sum\nolimits_{{\bf{a}}: \sim {a_i}} $ means the summation over all data symbols in $\bf{a}$ except $a_i$. The straightforward calculation of (\ref{app}) will involve complexity $O\left( {{2^N}} \right)$, which disastrously increases with $N$. 

For an efficient calculation of (\ref{app}) we will derive a factor graph representation for the probabilistic model of the system and employ the BP message passing algorithm onto the factor graph. To use the BP message passing algorithm and factor graphs, we first factorize the globe probability function $p\left( {\left. {\bf{Y}} \right|{\bf{a}}} \right)p\left( {\bf{a}} \right)$ into many small local functions. As specified by (\ref{dis_sample}), after the signal sampling of the proposed AcR, we meet an equivalent discrete memory channel, where the $i^{th}$ sample vector ${{\bf{Y}}_i}$ depends on the $M$ data symbols $\left\{ {{a_i},{a_{i - 1}}, \cdots ,{a_{i - M + 1}}} \right\}$. Therefore, we can factorize $p\left( {{\bf{Y}}\left| {\bf{a}} \right.} \right)$ as
\begin{equation}\label{cha_pro}
p\left( {{\bf{Y}}\left| {\bf{a}} \right.} \right) = \prod\limits_{i = 1}^N {p\left( {{{\bf{Y}}_i}\left| {{a_i},{a_{i - 1}}, \cdots ,{a_{i - M + 1}}} \right.} \right)},
\end{equation}
where
\begin{equation}\label{cha_pro1}
\begin{array}{l}
 p\left( {{{\bf{Y}}_i}\left| {{a_i},{a_{i - 1}}, \cdots ,{a_{i - M + 1}}} \right.} \right) \\
 \;\;\;\;\;\;\; \propto \prod\limits_{m = 1}^M {\exp \left( { - {{{{\left| {{Y_{i,m}} - \left( {\prod\nolimits_{z = i- m + 1}^i {{a_z}} } \right)N_f^2{E_g}} \right|}^2}} \mathord{\left/
 {\vphantom {{{{\left| {{Y_{i,m}} - \left( {\prod\nolimits_{z = m + 1}^i {{a_z}} } \right)N_f^2{E_g}} \right|}^2}} {\sigma _n^2}}} \right.
 \kern-\nulldelimiterspace} {\sigma _n^2}}} \right)}  \\
 \end{array}
\end{equation}
is obtained by using the Gaussian approximation on the discrete noise components. Then, we perform a factorization on $p\left( {\bf{a}} \right)$:
\begin{equation}\label{a_p}
\begin{array}{l}
 p\left( {\bf{a}} \right) = \prod\limits_{i = 1}^N {p\left( {{a_i}} \right)}  \\
\;\;\;\;\;\;\;\; = \prod\limits_{i = 1}^N {p\left( {{a_i},{a_{i - 1}}, \cdots ,{a_{i - M + 1}}\left| {{a_{i - 1}},{a_{i - 2}}, \cdots ,{a_{i - M}}} \right.} \right)}  \\
 \end{array}
\end{equation}
which intuitively brings out the Markov property introduced by the sampling mechanism of the proposed AcR. The first equality of (\ref{a_p}) is due to the independence of $a_i$ without considering the channel encoding\footnote{If considering the channel encoding, the independence of $a_i$ can be approximately achieved by the interleaving operation.}; the second equality of  (\ref{a_p}) is due to  
$$
\begin{array}{l}
p\left( {{a_i}} \right) = \frac{{p\left( {{a_i}} \right)\prod\limits_{j = i - M}^{i - 1} {p\left( {{a_j}} \right)} }}{{\prod\limits_{j = i - M}^{i - 1} {p\left( {{a_j}} \right)} }} = \frac{{\prod\limits_{j = i - M}^i {p\left( {{a_j}} \right)} }}{{\prod\limits_{j = i - M}^{i - 1} {p\left( {{a_j}} \right)} }} \\ 
= \frac{{p\left( {{a_i},{a_{i - 1}}, \cdots ,{a_{i - M + 1}},{a_{i - M}}} \right)}}{{p\left( {{a_{i - 1}},{a_{i - 2}}, \cdots ,{a_{i - M}}} \right)}} \\ 
= \frac{{p\left( {{a_i},{a_{i - 1}}, \cdots ,{a_{i - M + 1}},{a_{i - 1}},{a_{i - 2}}, \cdots ,{a_{i - M}}} \right)}}{{p\left( {{a_{i - 1}},{a_{i - 2}}, \cdots ,{a_{i - M}}} \right)}} \\ 
= p\left( {{a_i},{a_{i - 1}}, \cdots ,{a_{i - M + 1}}\left| {{a_{i - 1}},{a_{i - 2}}, \cdots ,{a_{i - M}}} \right.} \right). \\ 
\end{array}
$$

\begin{figure}[!t]
	\centering
	\includegraphics[width=3.5in]{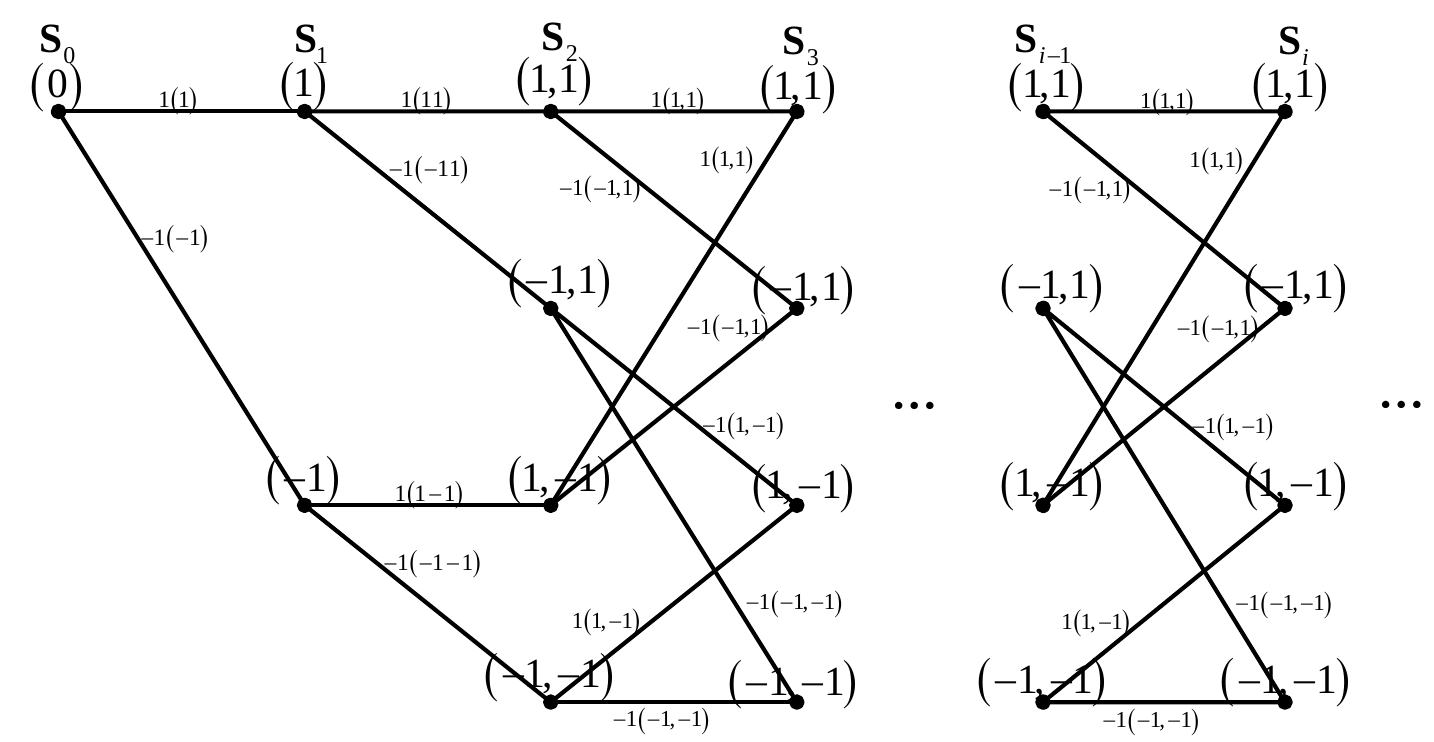}
	\caption{The trellis chart for the defined Makov chain with $M=2$. The digits one each transition represent the associated input(outputs): ${a_i}\left( {{{\bf{x}}_i}} \right)$.} \label{trellis}
\end{figure}

Based on the probability functions (\ref{cha_pro}) and (\ref{a_p}), we find that our system is well represented by a hidden Markov chain. We now  define such hidden Markov chain as follows. For the Markov chain, the $i^{th}$ state is ${{\bf{S}}_i}  \buildrel \Delta \over = {\left[ {{a_i},{a_{i - 1}}, \cdots ,{a_{i - M + 1}}} \right]^T}$, the $i^{th}$ input  is ${a_i}$ and the $i^{th}$ output is ${{\bf{x}}_i} \buildrel \Delta \over = {\left[ {{a_i},{a_{i - 1}}, \cdots ,{a_{i - M + 1}}} \right]^T}$. The behavior of the Markov chain is defined by the local check functions ${T_i}\left( {{a_i},{{\bf{x}}_i},{{\bf{S}}_{i-1}},{{\bf{S}}_{i }}} \right)$ that constrains the transitions from state $i-1$ to state $i$ to be valid. We have ${T_i}\left( {{a_i},{{\bf{x}}_i},{{\bf{S}}_{i-1}},{{\bf{S}}_{i }}} \right)=1$ when the combinations of its arguments are possible; ${T_i}\left( {{a_i},{{\bf{x}}_i},{{\bf{S}}_{i-1}},{{\bf{S}}_{i }}} \right)=0$ otherwise. The behavior of a Markov chain usually can be illustrated using a trellis chart. In Fig. \ref{trellis}, we give an example of such trellis chart that illustrates the Markov chain of our system with $M=2$.

\begin{figure*}[!t]
	%\normalsize
	\centering
	\includegraphics[width=7in]{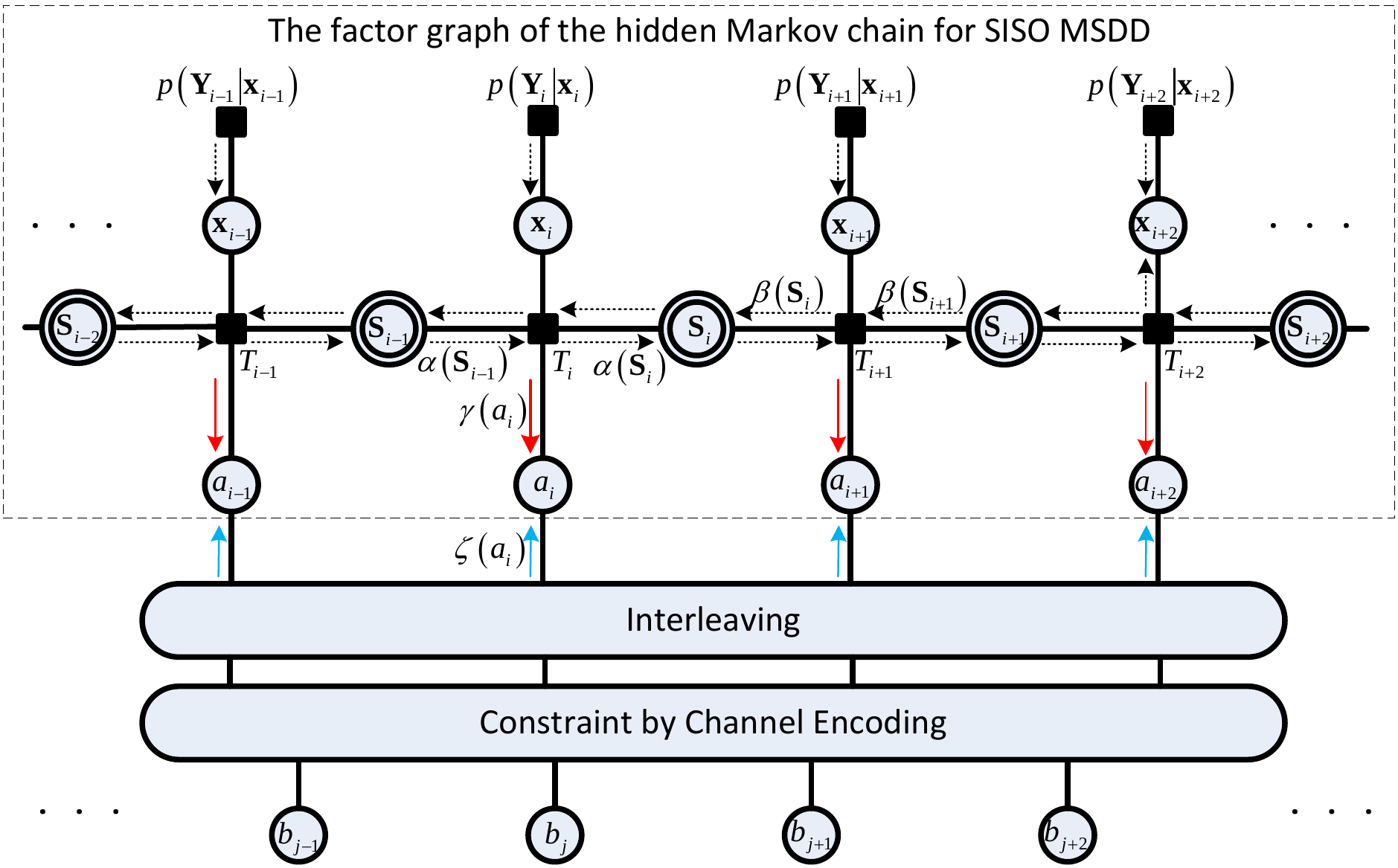}
	% IEEE uses as a separator
	%\hrulefill
	% The spacer can be tweaked to stop underfull vboxes.
	\caption{The factor graph for performing the proposed joint MSDD and channel decoding schme, where the factor graph in the dash line box is for performing the proposed SISO MSDD, the red arrows denote the intrinsic information sent from SISO MSDD to the channel decoding and the blue arrows denote the intrinsic information sent from the channel decoding to SISO MSDD.} \label{fg}
\end{figure*}

Our signal probabilistic model is a \emph{hidden} Markov chain due to that we cannot directly observe the output ${{\bf{x}}_i}$, we can only observe 
${{\bf{Y}}_i}$ which is a function of ${{\bf{x}}_i}$ plus a noise. Hidden Markov chains can be represented using factor graphs \cite{kschischang2001factor}. The factor graph representation for the hidden Markov chain of our signal probabilistic model is shown in the upper part of Fig. \ref{fg}, where circles are variable nodes for inputs and outputs, double circles are variable nodes for state and squares are the factor nodes for local check functions. Then, we can apply BP message passing algorithm to the factor graph for performing SISO MSDD. Since the structure of the factor graph for SISO MSDD is a tree, SISO MSDD is exactly implemented by a forward message
passing and a backward message passing. We give the detailed descriptions about these message passing operations in the following.

The above definition for the hidden Markov chain is consistent with the factorization in (\ref{a_p}): 
$$\begin{array}{l}
 p\left( {{a_i}} \right){T_i}\left( {{a_i},{{\bf{x}}_i},{{\bf{S}}_{i - 1}},{{\bf{S}}_i}} \right) \\ 
  = p\left( {{a_i},{a_{i - 1}}, \cdots ,{a_{i - M + 1}}\left| {{a_{i - 1}},{a_{i - 2}}, \cdots ,{a_{i - M}}} \right.} \right) \\ 
 \end{array}$$
Therefore, we have
 \begin{equation}\label{a_p_2}
p\left( {\bf{a}} \right) \propto\prod\limits_{i = 1}^N {p\left( {{a_i}} \right){T_i}\left( {{a_i},{{\bf{x}}_i},{{\bf{S}}_i},{{\bf{S}}_{i - 1}}} \right)}.
\end{equation}
Substituting (\ref{cha_pro}), (\ref{a_p_2}) into (\ref{app}) leads to
\begin{equation}\label{app2}
\begin{array}{*{20}{l}}
   {\begin{array}{*{20}{l}}
   {p\left( {\left. {{a_i}} \right|{\bf{Y}}} \right)}  \\
   { \propto \underbrace {\left( {{\sum _{{{\bf{a}}_{1:i - 1}}}}\prod\limits_{j = 1}^{i - 1} {p\left( {{a_j}} \right){T_j}\left( {{a_j},{{\bf{x}}_j},{{\bf{S}}_{j - 1}},{{\bf{S}}_j}} \right)p\left( {\left. {{{\bf{Y}}_j}} \right|{{\bf{x}}_j}} \right)} } \right)}_{ \buildrel \Delta \over = \alpha \left( {{{\bf{S}}_{i - 1}}} \right)}}  \\
\end{array}}  \\
   { \times \underbrace {\left( {{\sum _{{{\bf{a}}_{i + 1:N}}}}\prod\limits_{j = i + 1}^N {p\left( {{a_j}} \right){T_j}\left( {{a_j},{{\bf{x}}_j},{{\bf{S}}_{j - 1}},{{\bf{S}}_j}} \right)p\left( {\left. {{{\bf{Y}}_j}} \right|{{\bf{x}}_j}} \right)} } \right)}_{ \buildrel \Delta \over = \beta \left( {{{\bf{S}}_i}} \right)}}  \\
   { \times p\left( {{a_i}} \right){T_i}\left( {{a_i},{{\bf{x}}_i},{{\bf{S}}_{i - 1}},{{\bf{S}}_i}} \right)p\left( {{{\bf{Y}}_i}\left| {{{\bf{x}}_i}} \right.} \right)}.  \\
\end{array}
\end{equation}
As indicated in (\ref{app2}), we can now calculate APPs $p\left( {\left. {{a_i}} \right|{\bf{Y}}} \right)$ using a BP message passing algorithm that implements the sum-product rule \cite{kschischang2001factor}. The BP message passing algorithm for calculating $p\left( {\left. {{a_i}} \right|{\bf{Y}}} \right)$ is a 
bidirectional algorithm consisting of a forward and a backward
message passing. This is similar to the BCJR algorithm for decoding convolutional channel codes \cite{bahl1974optimal}.

Using the sum-product rule, the forward messsage passing aims to recursively calculate messages $\alpha \left( {{{\bf{S}}_i}} \right)$ (already defined in (\ref{app2})): 
\begin{equation}\label{forward}
\begin{array}{l}
 \alpha \left( {{{\bf{S}}_i}} \right) = \\ 
 \sum\limits_{\sim{{\bf{S}}_i}} {\alpha \left( {{{\bf{S}}_{i - 1}}} \right)p\left( {{a_i}} \right){T_i}\left( {{a_i},{{\bf{x}}_i},{{\bf{S}}_{i - 1}},{{\bf{S}}_i}} \right)p\left( {\left. {{{\bf{Y}}_i}} \right|{{\bf{x}}_i}} \right)}  \\ 
 \end{array}
\end{equation}
for each $i = 1,2, \cdots ,N$, where notaiton $\sum\nolimits_{ \sim {{\bf{S}}_i}} $ denotes the summation over all arguments involved in  except ${{\bf{S}}_i}$, $p\left( {{a_i}} \right)$ is the \emph{a priori} information of ${a_i}$, ${T_i}\left( {{a_i},{{\bf{x}}_i},{{\bf{S}}_{i - 1}},{{\bf{S}}_i}} \right)$ is the previously defined local check function for the state transition, and ${p\left( {\left. {{{\bf{Y}}_i}} \right|{{\bf{x}}_i}} \right)}$ is the evidence information from observation ${\bf{Y}}_i$ (shown in (\ref{cha_pro1})). Similarly, the backward messsage passing aims to recursively  calculate messages $\beta \left( {{{\bf{S}}_i}} \right)$ (already defined in (\ref{app2})):
\begin{equation}\label{backward}
\begin{array}{l}
 \begin{array}{*{20}{l}}
   {\beta \left( {{{\bf{S}}_i}} \right) = }  \\
   {\sum\limits_{\sim{{\bf{S}}_{i + 1}}} {\beta \left( {{{\bf{S}}_{i + 1}}} \right)p\left( {{a_{i + 1}}} \right){T_{i + 1}}\left( {{a_{i + 1}},{{\bf{x}}_{i + 1}},{{\bf{S}}_i},{{\bf{S}}_{i + 1}}} \right)} }  \\
\end{array} \\ 
  \;\;\;\;\;\;\;\;\;\;\;\;\;\;\;\;\;\;\;\;\;\;\;\;\;\;\;\;\;\;\;\;\;\;\;\;\;\;\;\;\;\;\;\;\;\;\;\;\;\;\;\;\; \times p\left( {\left. {{{\bf{Y}}_{i + 1}}} \right|{{\bf{x}}_{i + 1}}} \right) \\ 
 \end{array}
\end{equation}
for each $i = 1,2, \cdots ,N$. After the forward and backward message passing once in each direction, we can calculate the messages that runs fron check node $T_i$ to variable node $a_i$ (denoted by the red solid arrows in Fig. \ref{fg}) as
\begin{equation}\label{merge}
\begin{array}{*{20}{l}}
   {\gamma \left( {{a_i}} \right) = }  \\
   {{\sum _{\sim{a_i}}}\alpha \left( {{{\bf{S}}_{i - 1}}} \right)\beta \left( {{{\bf{S}}_i}} \right){T_i}\left( {{a_i},{{\bf{x}}_i},{{\bf{S}}_{i - 1}},{{\bf{S}}_i}} \right)p\left( {\left. {{{\bf{Y}}_i}} \right|{{\bf{x}}_i}} \right)}  \\
\end{array}
\end{equation}
for each $i = 1,2, \cdots ,N$. Finally, the APP of $a_i$ is given by 
$${p\left( {\left. {{a_i}} \right|{\bf{Y}}} \right) \propto p\left( {{a_i}} \right)\gamma \left( {{a_i}} \right)}$$
We now finish the derivations on the BP message passing algorithm for SISO MSDD.

\subsection{Joint Noncoherent Detection and Channel Decoding}
BP message passing algorithm is also widely used as the decoding algorithm for many advanced channel codes, such as LDPC codes, Turbo codes \cite{wiberg1995codes, kschischang2001factor}. It is straightforward to integrate BP for SISO MSDD with BP for channel decoding under the message passing framework, resulting in a BP message passing algorithm for joint noncoherent detection and channel decoding. In this subsection, we incorporate channel encoding/decoding into our framework. The factor graph of the overall system that includes the channel encoding constraint is shown in Fig. \ref{fg}.

The presence of the constraint on $\bf{a}$ by channel encoding introduces loops onto the factor graph. As a consequence, the BP message passing algorithm cannot exactly calculate these APPs of interest. On the factor graph withe loops, the BP message passing algroithm approximates the calculation of APPs in an iterative manner --- the messages will be passed multiple times on some given edges of the factor graph \cite{kschischang2001factor}. Usually, a good channel code is designed to make the loops very large. Therefore, in many applications, the approximations by iterative BP message passing on factor graphs with loops are pretty good \cite{kschischang2001factor}.

Given the factor graph of interest, we can design many different message-passing schedules. In this work, we adopt a serial schedule \cite{kschischang2001factor} for the iterative BP message passing between the SISO MSDD and the SISO channel decoding. The messages exchanged between the MSDD and the channel decoding are known as the extrinsic information. In each iteration, given the channel evidences $p\left( {{{\bf{Y}}_i}\left| {{{\bf{x}}_i}} \right.} \right)$ from observations ${\bf{Y}}$ and the messages $\zeta \left( {{a_i}} \right)$ sent from the channel decoding for all $i$, we performs the BP message passing algorithm for SISO MSDD to update messages $\gamma \left( {{a_i}} \right)$ as in (\ref{merge}) for all $i$. The update of $\gamma \left( {{a_i}} \right)$ is still according to (\ref{forward}), (\ref{backward}) and (\ref{merge}) with the only difference that we replace the \emph{a priori} information $p\left( {{a_i}} \right)$ in (\ref{forward}) and (\ref{backward}) with the  messages $\zeta \left( {{a_i}} \right)$. Then, the updated $\gamma \left( {{a_i}} \right)$ are treated as the extrinsic information and delivered to the channel decoding (denoted by the blue solid arrows in Fig. \ref{fg}).  Then, treating messages $\gamma \left( {{a_i}} \right)$ as the \emph{a priori} information, the channel decoding runs several rounds of iterative BP message passing within the subgraph of the channel encoding constraint. After that, the channel decoding sends back its extrinsic information $\zeta \left( {{a_i}} \right)$ (denoted by the blue dot arrows in Fig. \ref{fg}) to the MSDD for the next iteration. After several iterations between the MSDD and the channel decoding, we terminate the algorithm, and obtain the final decoding results about information bits. Finally, we remark that the above iterative processing is implemented in digital domain as long as we have obtained the correlation samples from the AcR receiver.

\subsection{Alternative SISO MSDD Scheme}

In Section III.B, we establish a hidden Markov chain for the signal model induced by the sampling mechanism of the proposed AcR; we then develope  a BP message passing algorithm on the factor graph of the hidden Markov chain for fulfilling SISO MSDD. Henceforth, we will refer to this hidden Markov chain based MSDD as the M-MSDD scheme. 

In \cite{guo2006improved, yang2008noncoherent, lottici2008multiple}, there is another kind of AcR proposed to generate correlation samples for MSDD. The sampling mechanism of the AcR and the following MSDD scheme in \cite{guo2006improved, yang2008noncoherent, lottici2008multiple} is in a block-by-block basis. We refer to this block based MSDD scheme as the B-MSDD scheme. Originally, the B-MSDD scheme proposed in \cite{guo2006improved, yang2008noncoherent, lottici2008multiple} only gives hard outputs. It can also be modified to become SISO B-MSDD and integrated with SISO channel decoding, as investigated in \cite{zhou2012soft}. In this subsection, we interpret the SISO B-MSDD scheme using the factor graph and BP message passing algorithm framework.

The whole packet of $N$ data symbols ${a_i}$, $i = 1, \cdots ,M$ is divided into $U = {N \mathord{\left/
 {\vphantom {N M}} \right.
 \kern-\nulldelimiterspace} M}$ blocks, where $M$ is the block size. The ${u^{th}}$ block, $u = 1, \cdots ,U$, includes $M$ data symbols ${a_i}$, $i = \left( {u - 1} \right)M + 1, \cdots ,uM$. We stack the symbols of the ${u^{th}}$ block into vector ${\widetilde{\bf{x}}_u} = {\left[ {{a_{\left( {u - 1} \right)M + 1}},{a_{\left( {u - 1} \right)M + 2}}, \cdots ,{a_{uM}}} \right]^T}$. For B-MSDD, the data symbols in the same block will be detected jointly. However, different blocks are processed independently. For the ${u^{th}}$ block, the AcR for B-MSDD samples the received signal within the observation window with size $M+1$ symbol durations, $r\left( t \right)$ for $\left( {u - 1} \right)M{T_s} \le t \le \left( {uM + 1} \right){T_s}$, using the following sampling mechanism
\begin{equation}\label{sampling2}
\begin{array}{*{20}{c}}
   \begin{array}{c}
 {\widetilde Y_{i,j}} = \int\limits_0^{{T_g}} {y\left( {t + i{T_s}} \right)y\left( {t + j{T_s}} \right)dt}  \\ 
 \;\;\;\;\; = \left( {\prod\limits_{z = j + 1}^i {{a_z}} } \right)N_f^2{E_g} + {n_{i,j}}, \\ 
 \end{array}  \\
   {\:\:\:\:\:\:\:\:\:\:\:\:\:\:\:\:\:\:\:\:\:\:\:\:\:\:\:\:\:\:\:\:\:\:\:i = \left( {u - 1} \right)M + 1, \cdots ,uM}  \\
   {\:\:\:\:\:\:\:\:\:\:\:\:\:\:\:\:\:\:\:\:\:\:\:\:\:\:\:\:\:\:\:\:\:\:\:j = \left( {u - 1} \right)M, \cdots ,i - 1}  \\
\end{array}
\end{equation}
where $y\left( t \right)$ is the de-spreading signal given in (\ref{despreading_signal}), ${n_{i,j}}$ is the discrete noise component. We stack all the samples ${\widetilde Y_{i,j}}$
of the ${u^{th}}$ block into a vector ${\widetilde{\bf{Y}}_u}$ of length $\sum\nolimits_{m = 1}^M m  = {{M\left( {M + 1} \right)} \mathord{\left/
 {\vphantom {{M\left( {M + 1} \right)} 2}} \right.
 \kern-\nulldelimiterspace} 2}$. Then, the observation window is slid down $M$ symbol durations for the sampling operations of the next $M$ symbols. Based on ${\widetilde{\bf{Y}}_u}$, B-MSDD proposed in \cite{guo2006improved, yang2008noncoherent, lottici2008multiple} finds hard decisons on the symbols in ${\widetilde{\bf{x}}_u}$ jointly.

\begin{figure}[!t]
\centering
\includegraphics[width=3.5in]{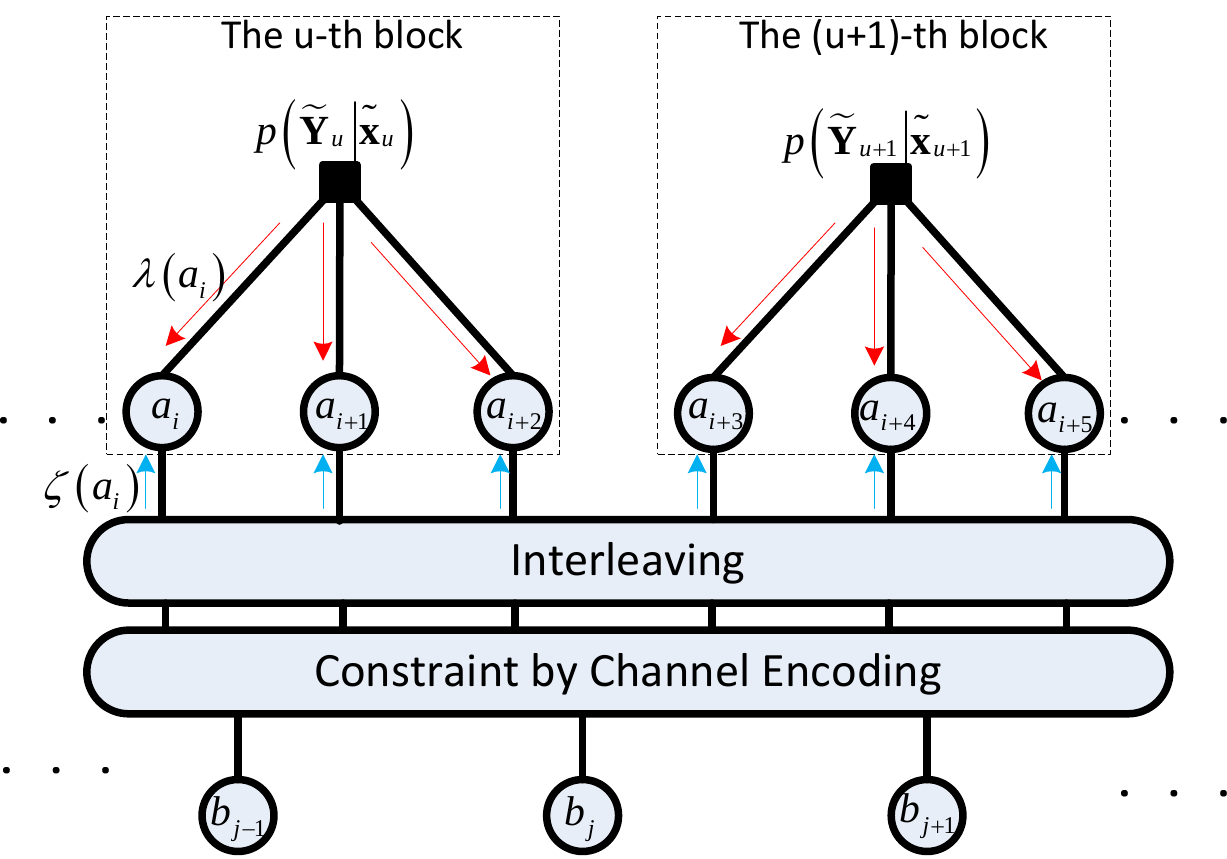}
\caption{The factor graph for performing joint B-MSDD and channel decoding with block size $M=3$, where $i = (u-1)M+1$, the red arrows denote the intrinsic information sent from SISO B-MSDD to the channel decoding and the blue arrows denote the intrinsic information sent from the channel decoding to SISO B-MSDD.}\label{fg2}
\end{figure}

For an SISO B-MSDD, we first construct the probabilistic model of the system using factor graphs. The factor graph for B-MSDD with $M=3$ as an example is shown in Fig.~\ref{fg2}, where we also incoporate the constraint by the channel encoding. Then, we develop a BP message passing algorithm on the factor graph. To be integrated with the SISO channel decoding, the SISO B-MSDD scheme aims to calculate messages $\lambda \left( {{a_i}} \right)$ (denoted by the red arrows in Fig.~\ref{fg2}):
\begin{equation}\label{bmsdd}
\lambda \left( {{a_i}} \right) = \sum\limits_{{{\bf{x}}_u}:\sim{a_i}} {p\left( {{{\widetilde{\bf{Y}}}_u}\left| {{{\widetilde{\bf{x}}}_u}} \right.} \right)\prod\limits_{j \in {I_u} \cap j \ne i} {\zeta \left( {{a_j}} \right)} } 
\end{equation}
for $i = \left( {u - 1} \right)M + 1, \cdots ,uM$, $u = 1, \cdots ,U$, where ${I_{u}} = \left\{ {\left( {u - 1} \right)M + 1, \cdots ,uM} \right\}$, the channel evidence information  
 $$\begin{array}{l}
  p\left( {{{\widetilde{\bf{Y}}}_u}\left| {{{\widetilde{\bf{x}}}_u}} \right.} \right) \\ 
   \propto \prod\limits_{i = \left( {u - 1} \right)M + 1}^{uM} {\prod\limits_{j = \left( {u - 1} \right)M}^{i - 1} {\exp \left( -{\frac{{ {{\left| {{Y_{i,j}} - \left( {\prod\limits_{z = j + 1}^i {{a_z}} } \right)N_f^2{E_g}} \right|}^2}}}{{\sigma _n^2}}} \right)} }  \\ 
  \end{array}$$
is obtained by using the Gaussian approximation on the discrete noise component, ${\zeta \left( {{a_j}} \right)}$ is the \emph{a priori} information of $a_j$ sent from the channel decoding (denoted by the blue arrows in Fig. \ref{fg2}). The messages $\lambda \left( {{a_i}} \right)$ are treated as the extrinsic inforamtion of B-MSDD and sent to the channel channel decoding. Then, treating messages $\lambda \left( {{a_i}} \right)$ as the \emph{a priori} information, the channel decoding updates its extrinsic information $\zeta \left( {{a_i}} \right)$ and send the updated $\zeta \left( {{a_i}} \right)$ back to B-MSDD. Several iterations between the B-MSDD and the channel decoding are performed to ensure the convergence of the joint B-MSDD and channel decoding algorithm.

The SISO B-MSDD scheme and the joint B-MSDD and channel decoding scheme are also investigated in \cite{zhou2012soft}. The extrinsic information of B-MSDD in \cite{zhou2012soft} is computed in the logarithm domain. The main contribution of \cite{zhou2012soft} is a practical solution to compute the logarithm-domain extrinsic information for a large block size $M$ using the list sphere decoding \cite{hochwald2003achieving}. Here, we just make an interpretation about SISO B-MSDD using the framework of the factor graph and the BP message passing algorithm. Thus, we compute the extrinsic information in the  probability domain using the sum-product rule as in (\ref{bmsdd}). The performance of B-MSDD is treated as the reference for our M-MSDD. We compare the performances of M-MSDD and B-MSDD through simulation study in the next section.

%\begin{figure*}[!t]
%\normalsize
%\includegraphics[width=3in]{factor_graph2}
%% IEEE uses as a separator
%%\hrulefill
%% The spacer can be tweaked to stop underfull vboxes.
%\caption{The factor graph for joint noncoherent detection and channel decoding, where the red arrows denote the intrinsic information sent from the BP-MSDD to the BP channel decoding and the blue arrows denote the intrinsic information sent from the BP channel decoding to the BP-MSDD.} \label{fg}
%\end{figure*}

\section{Simulation Results}

In this section, simulations are conducted to validate the proposed scheme. In all simulations, the channel are generated according to IEEE 802.15.3a CM2 model \cite{channel}, and the channel impulse responses are truncated at $T_g = 100$ ns. The used impulse shape $\omega(t)$  is the second derivative of a Gaussian function. The duration of $\omega(t)$ is set as $T_{\omega}=0.5$ ns. The frame and chip duration are set to $T_f = 200$ ns and $T_c = 1.0$ ns, respectively. Each symbol consists of $N_f = 10$ frames. The TH codes are randomly picked up in the interval $\left[ {0,{N_c} - 1} \right]$ where $N_c = 100$. Since now ${{\rm{T}}_f} > {T_g} + \left( {{N_c} - 1} \right){T_c}$ is satisfied, there is no IFI in our system.
The integration interval of AcR is $T_i=T_g=100$ ns. The bandwidth of the baseband filter employed at the receiver is 2 GHz. The signal-to-noise ration (SNR) is defined as ${{{N_f}{E_g}} \mathord{\left/
		{\vphantom {{{N_f}{E_g}} {\left( {R{N_0}} \right)}}} \right.
		\kern-\nulldelimiterspace} {\left( {R{N_0}} \right)}}$
\\

\begin{figure}[!t]
	\centering
	\includegraphics[width=3.4in]{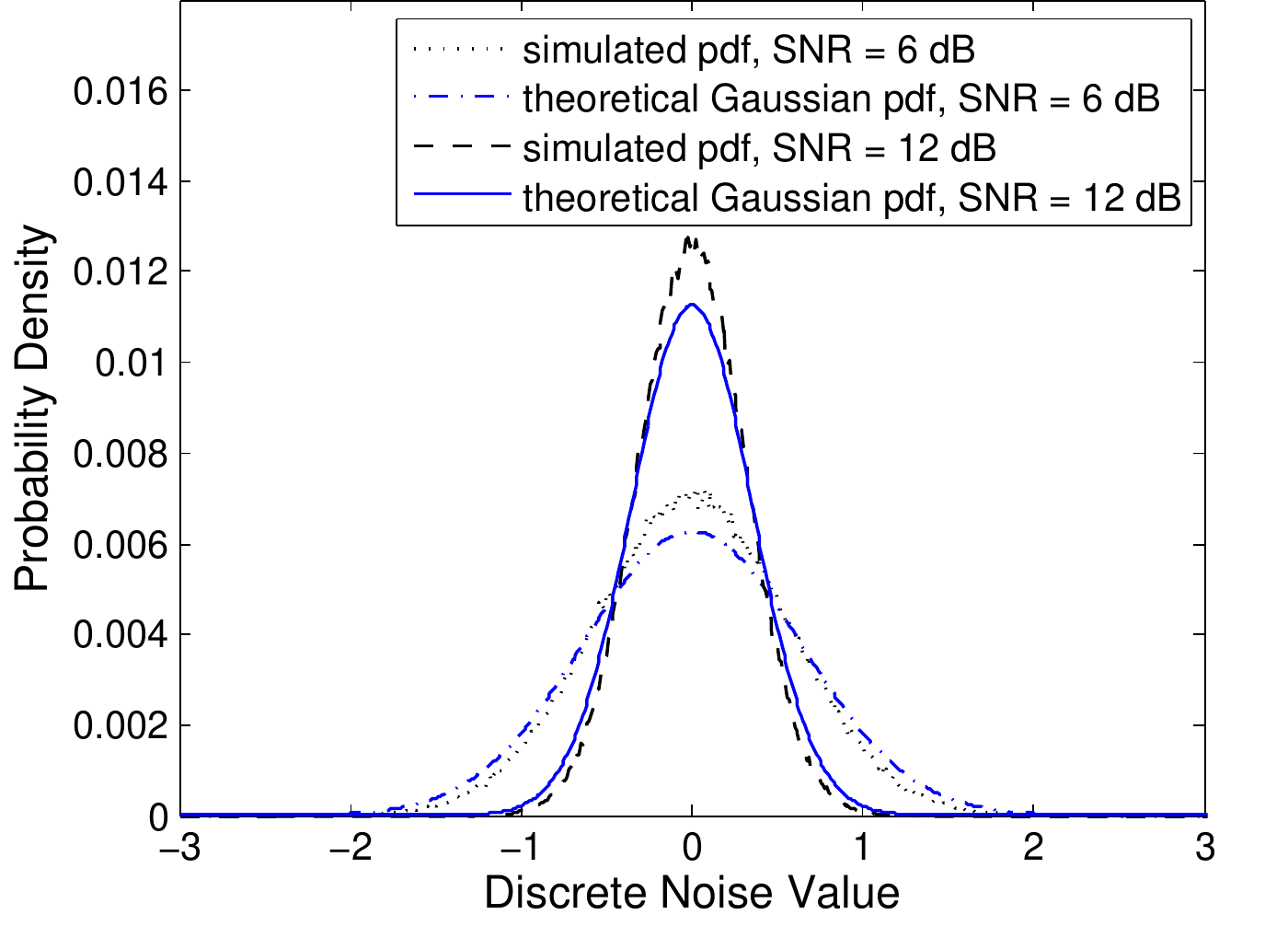}
	\caption{Comparison of the empirical noise PDFs to the theoretical Gaussian PDFs. The block sampling size is $M=3$.}\label{gauss}
\end{figure}

\noindent {\bf{\emph{Test Case 1}}}: 

This case is used to investigate the Gaussian approximation on the discrete noise components given in (\ref{dis_noise}). We get the correlation samples from the UWB signal waveforms using the sampling mechanism in (\ref{sampling}) with $M=3$. Then, we subtract $\left( {\prod\nolimits_{z =i- m + 1}^i {{a_z}} } \right) N_f^2{E_g}$ from ${Y_{i,m}}$ to get the noise component ${n_{i,m}}$ for all $m$ and $i$. We generate many noise components from $10^4$ pakects, each includes $N=1600$ data symbols (thus, $N_f(N+1)$ pulses). Finally, we compute the empirical noise probability density function (PDF) using these simulated noise components. The results are shown in Fig.~\ref{gauss}. The theoretical Gaussain PDF with mean zero and variance $\sigma _n^2 = {N_f}{N_0}{E_g} + {{W{T_g}N_0^2} \mathord{\left/{\vphantom {{W{T_g}N_0^2} 2}} \right. \kern-\nulldelimiterspace} 2}$ are also shown in Fig.~\ref{gauss} for comparison. From the results in Fig.~\ref{gauss}, we can see that the empirical noise PDFs can match the theoretical Gaussian PDFs.  
\\

\begin{figure}[!t]
	\centering
	\includegraphics[width=3.4in]{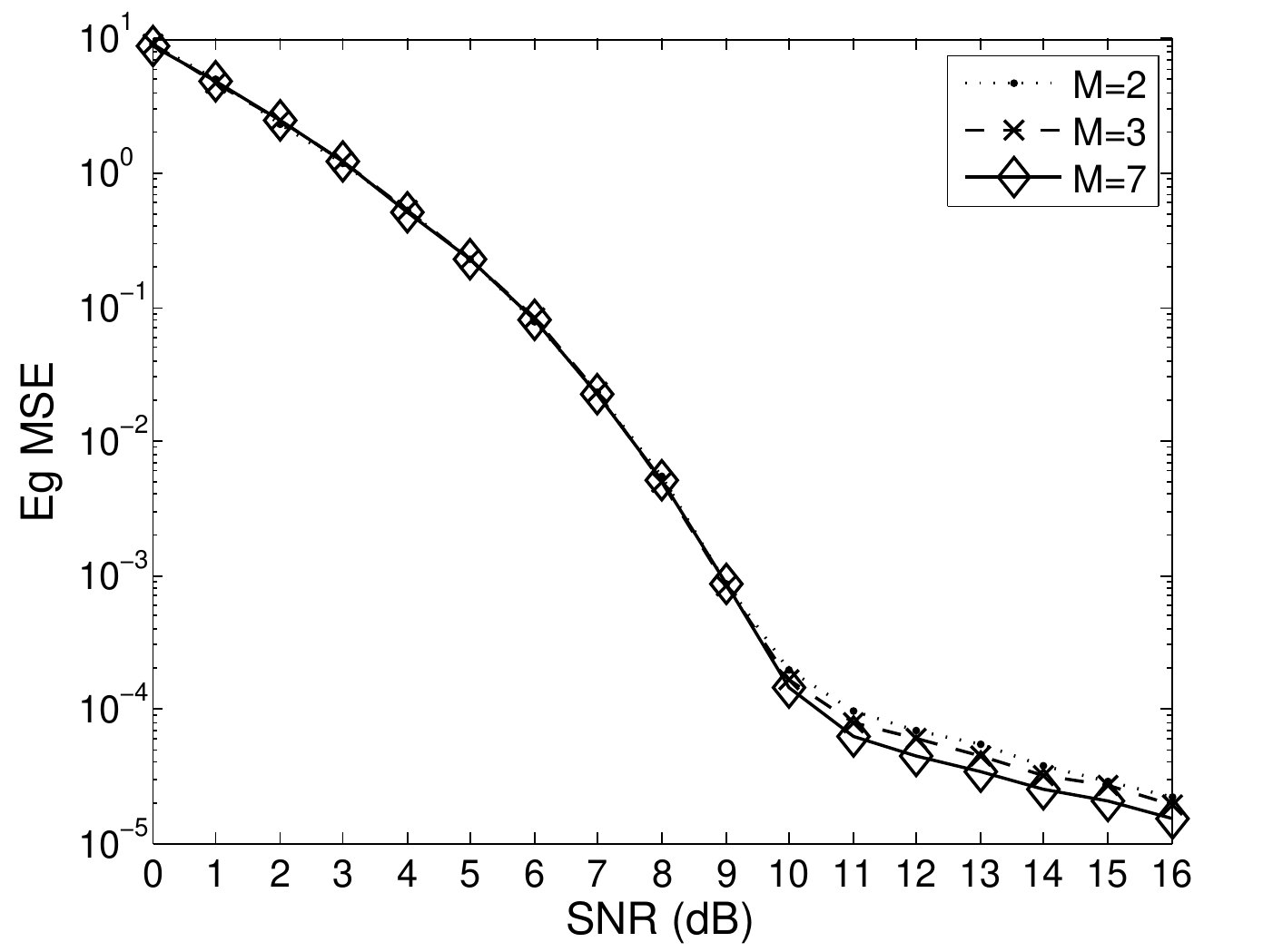}
	\caption{The MSE results of $E_g$ estimates. The used estimation method is given in (\ref{Eg}).} \label{mse}
\end{figure}

\noindent {\bf{\emph{Test Case 2}}}: 

We study whether the used estimation of $E_g$ given in (\ref{Eg}) works well. We compute the mean square error (MSE) of the $E_g$ estimates. We set that the packet includes $N=1600$ data symbols, and the block size used to sample the UWB signal waveform as in (\ref{sampling}) is $M = 2, 3, 7$. For each $M$ and each SNR, we average the square errors of $E_g$ estimates over $10^4$ packets to get the MSE. The MSE results are presented in Fig.~\ref{mse}. It shows that the values of MSE are small for the median to high SNR regime (above 7 dB) and stable for different $M$. In the high SNR regime (above 10 dB), we just see a little difference in MSE for different $M$; a larger $M$ induces a smaller MSE. This is because we can get more samples from lager $M$ and we can average out the noises better. Since the MSEs are relatively small, we will see later that the use of estimated $E_g$ in our scheme just introduce a negligible performance loss.       
\\

\noindent {\bf{\emph{Test Case 3}}}: 

We now investigate the bit error rate (BER) performance of the proposed M-MSDD scheme for uncoded differential UWB-IR systems. Without considering channel codes, we employ the M-MSDD to detect the data symbols ${a_i}$ for $i = 1,2, \cdots ,N$. After the bidirectional message passing as in (\ref{forward}), (\ref{backward}) and (\ref{merge}), the BP message passing algorithm for M-MSDD outputs the APP of ${a_i}$. Then, the hard decision about ${a_i}$ is made based upon the APP of ${a_i}$. Each packet consists of $N=1600$ data symbols. We evaluate the BER performance of the proposed M-MSDD scheme with the perfect $E_g$ and the estimated $E_g$. As benchmarks, we also evaluate the BER performances of the DD \cite{ho2002differential} scheme and the hard B-MSDD \cite{guo2006improved, yang2008noncoherent, lottici2008multiple} scheme for differential UWB-IR systems.

\begin{figure}[!t]
	\centering
	\includegraphics[width=3.4in]{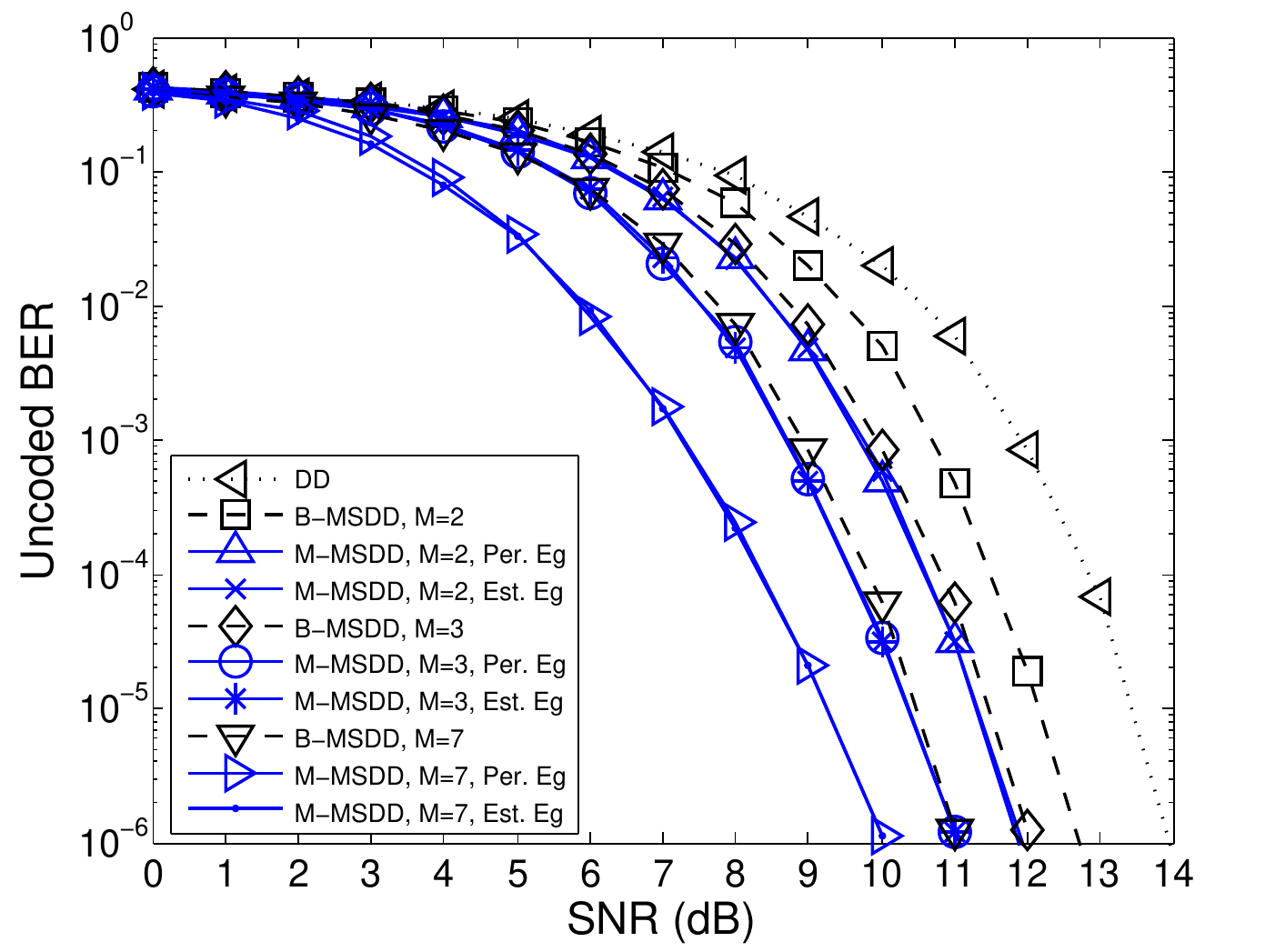}
	\caption{The BER results of the uncoded system. } \label{uncoded_ber}
\end{figure}

Fig.~\ref{uncoded_ber} presents the BER results. The first point we want to study is the impact of $E_g$ on the BER performance of our M-MSDD. We can observe from Fig.~\ref{uncoded_ber} that the M-MSDD schemes with perfect $E_g$ and estimated $E_g$ nearly have the same performance. Thus, we can conclude that the estimate of $E_g$ by the simple energy estimation method is sufficiently effective for the implementation of M-MSDD. Then, compared to the DD scheme, our M-MSDD scheme can improve the uncoded BER performance by offering detection gains (2, 3, 4 dB at the BER of $10^{-6}$ for $M=2,3,7$) which increase with $M$. This performance trend in improving BER by M-MSDD is similar to that by B-MSDD. Since for a fixed $M$, M-MSDD and B-MSDD both have complexity in the order of $O(2^{M})$, we compare the uncoded BER performances of M-MSDD and B-MSDD with the same block size $M$. We see that M-MSDD has a better uncoded BER performance than B-MSDD --- M-MSDD obtains about 1 dB gain at the BER of $10^{-6}$ for a fixed block size $M$.
\\

\noindent {\bf{\emph{Test Case 4}}}: 

We then investigate the BER performances of the joint MSDD and channel decoding schemes for coded differential UWB-IR systems. The LDPC code \cite{mackay1999good} with coding rate $R=1/2$ is employed. Each packet has $800$ information bits (thus $1600$ channel-coded data symbols). In the joint MSDD and channel decoding schemes, the used MSDD schemes are our SISO M-MSDD proposed in Section III.B and the SISO B-MSDD discussed in Section III.D, respectively.  We evaluate the BER performance of the schemes with the perfect $E_g$ and the estimated $E_g$. For all simulation results, we perform $10$ iterations between the BP message passing algorithm for MSDD and the BP message passing algorithm for LDPC channel decoding, and $10$ iterations within the BP message passing algorithm for LDPC channel decoding. In all simulations, we observe that these numbers of iterations can ensure the convergence of the algorithms.

\begin{figure}[!t]
	\centering
	\includegraphics[width=3.4in]{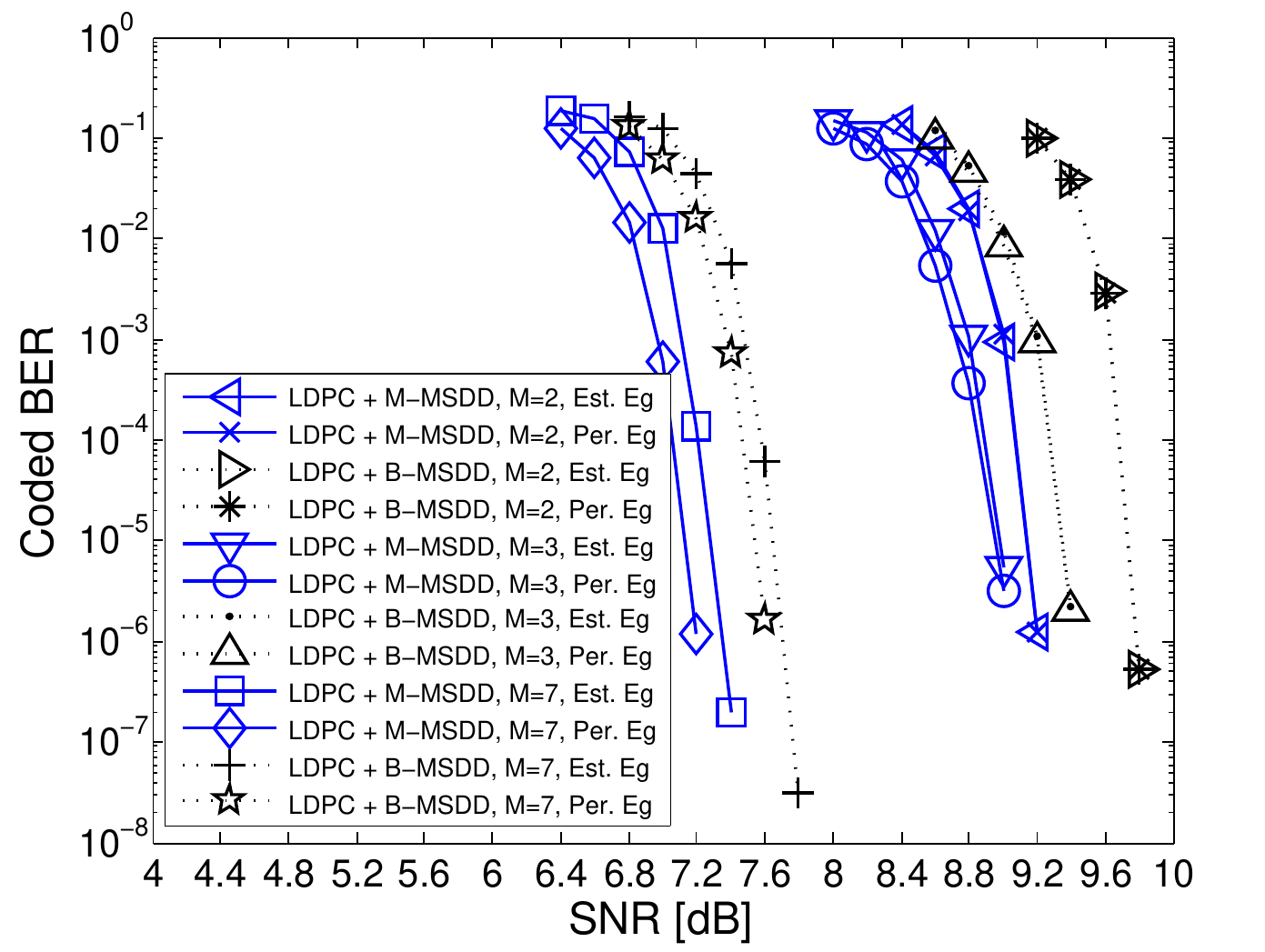}
	\caption{The BER results of the coded system.} \label{coded_ber}
\end{figure}

Fig.~\ref{coded_ber} presents the coded BER results. The first observation is that for the schemes with a large block size $M$, the estimated $E_g$ now induces some performance loss in the coded BER. The reason is that the coded system with a large $M$ is operating at a relatively low SNR regime, where the estimation error of $E_g$ is large and the decoding performance is sensitive to the estimation error. Second, we can see that the proposed joint M-MSDD and channel decoding scheme has around $0.4$ dB SNR gain over the joint B-MSDD and channel decoding scheme for a fixed block size $M$. We believe that these gains are due to the more beliefs collected by the BP message passing algorithm for M-MSDD.

\begin{figure}[!t]
	\centering
	\includegraphics[width=3.4in]{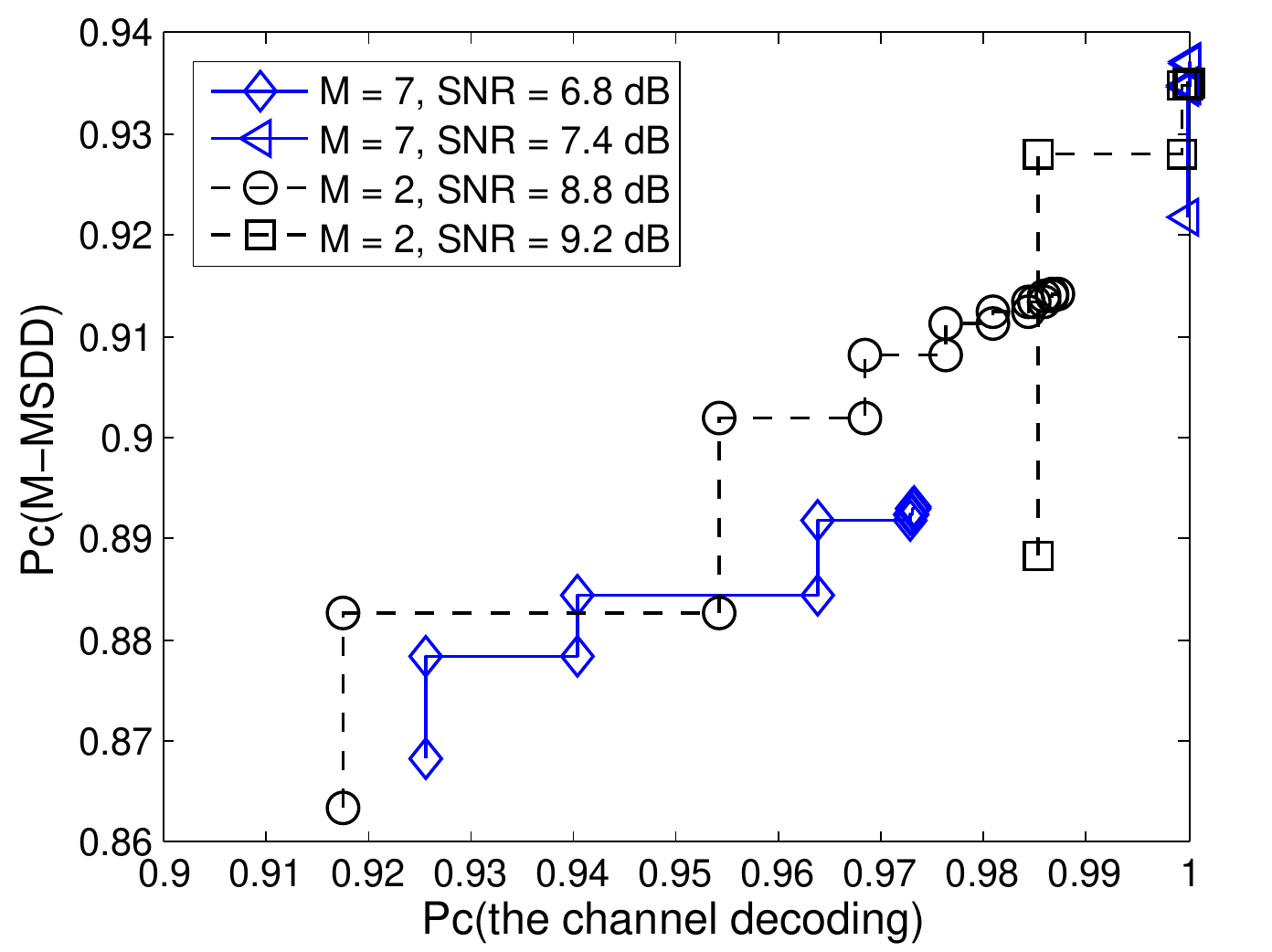}
	\caption{Probability of symbol decisions through iterations.} \label{ber_chart}
\end{figure}

In order to show the convergence of the iterative process between M-MSDD and the channel decoding, we track the correct probability of the symbol decisions for $a_i$, $i = 1,2, \cdots ,N$ through iterations. The results are shown in Fig.~\ref{ber_chart}. The $x$ axis represents ${P_c}\left( {{\rm{the~ channel~ decoding}}} \right)$ , the correct probability of the symbol decisions at the output of the channel decoding; the $y$ axis represents ${P_c}\left( {{\rm{M - MSDD}}} \right)$, the correct probability of the symbol decisions at the output of M-MSDD. The decisions are made based on the extrinsic information at the outputs. The coordinates of each point indicate the correct probability of the symbol decisions at the output of M-MSDD and the channel decoding at the end of a certain iteration. From the results, we find that at a lager SNR operating point, the algorithm have a faster convergence. For example, when SNR = 8.8 dB, the algorithm with $M=2$ has converged nearly after 8 iterations between M-MSDD and the channel decoding; when SNR = 9.2 dB, the algorithm with $M=2$ has converged nearly after 3 iterations.

\section{Conclusion}

In this paper, we apply BP message passing algorithm to propose a joint MSDD and channel decoding scheme for noncoherent UWB-IR systems. Specifically, we propose a new AcR architecture to transform the received UWB-IR signal into discrete samples, whose probabilistic model has a hidden Markov chain structure. Using the factor graph representation of this hidden Markov chain model and applying the BP message passing algorithm on the factor graph, we derive a new SISO MSDD for computing the APPs of the data symbols. The proposed MSDD is a bidirectional message passing algorithm, which can make use of all the signal dependences throughout the whole packet. Thus, the proposed MSDD has a better performance than the block-by-block MSDD scheme proposed in previous works. We can also feed the outputs of MSDD to the inputs of the BP message passing algorithm for channel decoding, and vice verse, in an iterative manner. Our simulations indicate that the proposed MSDD scheme has 1 dB, and 0.4 dB gain over the block-by-block MSDD scheme for the uncoded and coded system, respectively.

%\appendix
%
%
%The first equality of (\ref{a_p}) is straightforward. We now prove the second equality of  (\ref{a_p}). Using the conditional probability formula, we can write
%$$
%\begin{array}{l}
%	p\left( {{a_i},{a_{i - 1}}, \cdots ,{a_{i - M + 1}}\left| {{a_{i - 1}},{a_{i - 2}}, \cdots ,{a_{i - M}}} \right.} \right) \\ 
%	= \frac{{p\left( {{a_i},{a_{i - 1}}, \cdots ,{a_{i - M + 1}},{a_{i - 1}},{a_{i - 2}}, \cdots ,{a_{i - M}}} \right)}}{{p\left( {{a_{i - 1}},{a_{i - 2}}, \cdots ,{a_{i - M}}} \right)}} \\ 
%	= \frac{{p\left( {{a_i},{a_{i - 1}}, \cdots ,{a_{i - M + 1}},{a_{i - M}}} \right)}}{{p\left( {{a_{i - 1}},{a_{i - 2}}, \cdots ,{a_{i - M}}} \right)}} \\ 
%	= \frac{{\prod\limits_{j = i - M}^i {p\left( {{a_j}} \right)} }}{{\prod\limits_{j = i - M}^{i - 1} {p\left( {{a_j}} \right)} }} \\ 
%	= p\left( {{a_i}} \right) \\ 
%\end{array}
%$$
%which finish the proof.

\ifCLASSOPTIONcaptionsoff
  \newpage
\fi

\bibliographystyle{IEEEtran}
\bibliography{symbollevelcombining}

% Generated by IEEEtran.bst, version: 1.13 (2008/09/30)
\begin{thebibliography}{10}
\providecommand{\url}[1]{#1}
\csname url@samestyle\endcsname
\providecommand{\newblock}{\relax}
\providecommand{\bibinfo}[2]{#2}
\providecommand{\BIBentrySTDinterwordspacing}{\spaceskip=0pt\relax}
\providecommand{\BIBentryALTinterwordstretchfactor}{4}
\providecommand{\BIBentryALTinterwordspacing}{\spaceskip=\fontdimen2\font plus
\BIBentryALTinterwordstretchfactor\fontdimen3\font minus
  \fontdimen4\font\relax}
\providecommand{\BIBforeignlanguage}[2]{{%
\expandafter\ifx\csname l@#1\endcsname\relax
\typeout{** WARNING: IEEEtran.bst: No hyphenation pattern has been}%
\typeout{** loaded for the language `#1'. Using the pattern for}%
\typeout{** the default language instead.}%
\else
\language=\csname l@#1\endcsname
\fi
#2}}
\providecommand{\BIBdecl}{\relax}
\BIBdecl

\bibitem{yang2004uwc}
L.~Yang and G.~Giannakis, ``{Ultra-wideband communications: an idea whose time
  has come},'' \emph{IEEE Signal Process. Mag.}, vol.~21, no.~6, pp. 26--54,
  Jun. 2004.

\bibitem{rajeswaran2003rake}
A.~Rajeswaran, V.~S. Somayazulu, and J.~R. Foerster, ``Rake performance for a
  pulse based {UWB} system in a realistic {UWB} indoor channel,'' in
  \emph{Proc. 2003 IEEE Int. Conf. Commun.}, 2003, pp. 2879--2883.

\bibitem{win1998ecu}
M.~Win and R.~Scholtz, ``{On the energy capture of ultrawide bandwidth signals
  in dense multipath environments},'' \emph{IEEE Commun. Lett.}, vol.~2, no.~9,
  pp. 245--247, Sep. 1998.

\bibitem{lottici2002channelestimation}
V.~Lottici, A.~D'Andrea, and U.~Mengali, ``{Channel estimation for
  ultra-wideband communications},'' \emph{IEEE J. Sel. Areas Commun.}, vol.~20,
  no.~9, pp. 1638--1645, Sep. 2002.

\bibitem{chen2006timing}
J.~Chen, T.~Lv, Y.~Chen, and J.~Lv, ``A timing-jitter robust uwb modulation
  scheme,'' \emph{IEEE Signal Process. Lett.}, vol.~13, no.~10, pp. 593--596,
  2006.

\bibitem{witrisal2009noncoherent}
K.~Witrisal, G.~Leus, G.~Janssen, M.~Pausini, F.~Troesch, T.~Zasowski, and
  J.~Romme, ``{Noncoherent ultra-wideband systems},'' \emph{IEEE Signal
  Process. Mag.}, vol.~26, no.~4, pp. 48--66, Apr. 2009.

\bibitem{ho2002differential}
M.~Ho, V.~Somayazulu, J.~Foerster, and S.~Roy, ``{A differential detector for
  an ultra-wideband communications system},'' in \emph{Proc. IEEE 55th
  Vehicular Technology Conference}, May 2002, pp. 1896--1900.

\bibitem{choi2002performance}
J.~Choi and W.~Stark, ``{Performance of ultra-wideband communications with
  suboptimal receivers in multipath channels},'' \emph{IEEE J. Sel. Areas
  Commun.}, vol.~20, no.~9, pp. 1754--1766, Sep. 2002.

\bibitem{divsalar1990multiple}
D.~Divsalar and M.~Simon, ``{Multiple-symbol differential detection of MPSK},''
  \emph{IEEE Trans. Commun.}, vol.~38, no.~3, pp. 300--308, Mar. 1990.

\bibitem{guo2006improved}
N.~Guo and R.~Qiu, ``{Improved autocorrelation demodulation receivers based on
  multiple-symbol detection for UWB communications},'' \emph{IEEE Trans.
  Wireless Commun.}, vol.~5, no.~8, pp. 2026--2031, Aug. 2006.

\bibitem{yang2008noncoherent}
Y.~Tian and C.~Yang, ``Noncoherent multiple-symbol detection in coded
  ultra-wideband communications,'' \emph{IEEE Trans. Wireless Commun.}, vol.~7,
  no.~6, pp. 2202--2211, Jun. 2008.

\bibitem{lottici2008multiple}
V.~Lottici and Z.~Tian, ``{Multiple symbol differential detection for UWB
  communications},'' \emph{IEEE Trans. Wireless Commun.}, vol.~7, no.~5, pp.
  1656--1666, May 2008.

\bibitem{wang2011sphere}
T.~Wang, T.~Lv, and H.~Gao, ``Sphere decoding based multiple symbol detection
  for differential space-time block coded ultra-wideband systems,'' \emph{IEEE
  Commun. Lett.}, vol.~15, no.~3, pp. 269--271, Mar. 2011.

\bibitem{qi2010fast}
Q.~Zhou, X.~Ma, and V.~Lottici, ``Fast multi-symbol based iterative detectors
  for {UWB} communications,'' \emph{EURASIP Journal on Advances in Signal
  Processing}, vol. 2010, pp. 1--14, May 2010.

\bibitem{schenk2011decision}
A.~Schenk and R.~Fischer, ``Decision-feedback differential detection in
  impulse-radio ultra-wideband systems,'' \emph{IEEE Trans. Commun.}, vol.~59,
  no.~6, pp. 1604--1611, Jun. 2011.

\bibitem{wang2013ber}
T.~Wang, T.~Lv, H.~Gao, and Y.~Lu, ``{BER} analysis of decision-feedback
  multiple-symbol detection in noncoherent {MIMO} ultrawideband systems,''
  \emph{IEEE Trans. Veh. Technol.}, vol.~62, no.~9, pp. 4684 -- 4690, Nov.
  2013.

\bibitem{wiberg1995codes}
N.~Wiberg, H.-A. Loeliger, and R.~Kotter, ``Codes and iterative decoding on
  general graphs,'' \emph{European Trans. Telecommun.}, vol.~6, no.~5, pp.
  513--525, 1995.

\bibitem{zhou2012soft}
Q.~Zhou and X.~Ma, ``Soft-input soft-output multiple symbol differential
  detection for uwb communications,'' \emph{IEEE Commun. Lett.}, vol.~16,
  no.~8, pp. 1296--1299, 2012.

\bibitem{kschischang2001factor}
F.~R. Kschischang, B.~J. Frey, and H.-A. Loeliger, ``Factor graphs and the
  sum-product algorithm,'' \emph{IEEE Trans. Inf. Theory}, vol.~47, no.~2, pp.
  498--519, 2001.

\bibitem{molisch2006comprehensive}
A.~Molisch, D.~Cassioli, C.~Chong, S.~Emami, A.~Fort, B.~Kannan, J.~Karedal,
  J.~Kunisch, H.~Schantz, K.~Siwiak \emph{et~al.}, ``{A comprehensive
  standardized model for ultrawideband propagation channels},'' \emph{IEEE
  Trans. Antennas Propag.}, vol.~54, no.~11, p. 3151, November 2006.

\bibitem{quek2005analysis}
T.~Quek and M.~Win, ``Analysis of {UWB} transmitted-reference communication
  systems in dense multipath channels,'' \emph{IEEE J. Sel. Areas Commun.},
  vol.~23, no.~9, pp. 1863--1874, Sep. 2005.

\bibitem{bahl1974optimal}
L.~Bahl, J.~Cocke, F.~Jelinek, and J.~Raviv, ``Optimal decoding of linear codes
  for minimizing symbol error rate,'' \emph{IEEE Trans. Inf. Theory}, vol.~20,
  no.~2, pp. 284--287, 1974.

\bibitem{hochwald2003achieving}
B.~M. Hochwald and S.~Ten~Brink, ``Achieving near-capacity on a
  multiple-antenna channel,'' \emph{IEEE Trans. Commun.}, vol.~51, no.~3, pp.
  389--399, 2003.

\bibitem{channel}
J.~Foerster, ``{Channel Modeling Subcommittee Report Final (doc.:
  IEEE802-15-02/490rl-SG3a). IEEE P802. 15 Working Group for Wireless Personal
  Area Networks (WPANs), Feb. 2002}.''

\bibitem{mackay1999good}
D.~J. MacKay, ``Good error-correcting codes based on very sparse matrices,''
  \emph{IEEE Trans. Inf. Theory}, vol.~45, no.~2, pp. 399--431, 1999.

\end{thebibliography}

%\begin{figure}[!h]
%\centering
%\includegraphics[width=3.6in]{Fig1}
%% where an .eps filename suffix will be assumed under latex,
%% and a .pdf suffix will be assumed for pdflatex; or what has been declared
%% via \DeclareGraphicsExtensions.
%\caption{The BER performance of suboptimal ML-MSD for DSTBC-UWB
%system.} \label{fig_sim1}
%\end{figure}

% that's all folks
\end{document}